\documentclass[twocolumn,showpacs,preprintnumbers,amsmath,amssymb]{revtex4}

\usepackage{epsfig}
\usepackage{graphics}
\usepackage{graphicx}
\usepackage{dcolumn}
\usepackage{bm}

\begin{document}



\title{Zero-lag synchronization in coupled time-delayed piecewise linear electronic circuits}

\author{R.~Suresh$^1$}
\author{K.~Srinivasan$^1$}
\author{D.~V.~Senthilkumar$^{2}$}
\author{I. Raja Mohamed$^{3}$}
\author{K.~Murali$^4$}
\author{M.~Lakshmanan$^1$}
\author{J.~Kurths$^{2,5,6}$}

\affiliation{$^1$Centre for Nonlinear Dynamics, School of Physics, Bharathidasan University, Tiruchirapalli 620 024, India\\
$^2$Potsdam Institute for Climate Impact Research, 14473 Potsdam, Germany\\
$^3$Department of Physics, B. S. Abdur Rahman University, Chennai 600 048, India\\
$^4$Department of Physics, Anna University, Chennai 600 025, India\\
$^5$Institute of Physics, Humboldt University, 12489 Berlin, Germany\\
$^6$Institute for Complex Systems and Mathematical Biology, University of Aberdeen, Aberdeen AB24 3UE, United Kingdom}
\date{\today}

\begin{abstract}
We investigate and report an experimental confirmation of zero-lag synchronization (ZLS) in a system of three coupled
time-delayed piecewise linear electronic circuits via dynamical relaying with different 
coupling configurations, namely mutual and subsystem coupling configurations. 
We have observed that when there is a feedback between the central unit (relay unit) 
and at least one of the outer units, ZLS occurs in the two outer units whereas the central 
and outer units exhibit inverse phase synchronization (IPS).
 We find that in the case of mutual coupling configuration ZLS occurs both in periodic and 
hyperchaotic regimes, while in the subsystem coupling configuration it occurs only in the hyperchaotic
regime. Snapshots of the time evolution of outer circuits as observed from the oscilloscope confirm the
occurrence of ZLS experimentally. The quality of ZLS is numerically verified
by correlation coefficient and similarity function measures. 
Further, the transition to ZLS is verified
from the changes in the largest Lyapunov exponents and the correlation coefficient as a function of 
the coupling strength. IPS is experimentally confirmed 
using time series plots and also can be visualized using the concept of localized sets 
which are also corroborated by numerical simulations.
In addition, we have calculated the correlation of probability of recurrence to 
quantify the phase coherence.
We have also analytically derived a sufficient condition 
for the stability of ZLS using the Krasovskii-Lyapunov theory. 
\end{abstract}

\maketitle

{\bf Nonlinear time-delay systems constitute an important class of dynamical
systems which are abundant in nature and in real life. Synchronizing such systems is
having important implications in all areas of science. Different types of synchronization
have been numerically identified in a variety of ensembles of time-delay systems.
In contrast, only a very limited number of studies is available from experimental point of view. 
In this paper, we report a concrete experimental evidence of zero-lag synchronization (ZLS) 
in a system of three coupled time-delayed nonlinear 
electronic circuit with two different coupling configurations, namely mutual 
and subsystem coupling configurations. 
We have observed that when there is a feedback between the central and atleast one of the
outer units, ZLS occurs in the two outer units whereas central and outer units exhibit inverse phase 
synchronization (IPS). The results are experimentally confirmed using snap shots of the time evolution,
phase projection plots and the concept of localized sets along with corresponding numerical
results. The transition to ZLS and IPS can be quantified from the changes in the largest
Lyapunov exponents, correlation coefficient and correlation of probability of recurrence as a 
function of the coupling strength. We have also analytically deduced sufficient stability conditions
to confirm ZLS using the Krasovskii-Lyapunov theory.}

\section{\label{sec:level1}Introduction}
In 2006, Fischer et al. reported zero-lag synchronization (ZLS) in two mutually delay 
coupled chaotic units via a relay unit (also known as isochronal synchronization) \cite{if2006},
in the absence of which, both systems exhibit lead or lag synchronization. 
Interestingly, this phenomenon  could explain the occurrence of identical synchronization between widely
separated cortical regions of the human brain despite of synaptic/dendritic delays \cite{ake1991, prr1997, dc2001, rv2008}. 
ZLS has also been experimentally demonstrated
in delay coupled semiconductor lasers \cite{ae2010}, optoelectronic oscillators \cite{li2011} 
and in low-dimensional delay coupled chaotic electronic circuits (without intrinsic time-delay) \cite{aw2007} via dynamical relaying. 
Recently Banerjee et al. reported that the coupling threshold for ZLS 
between the outermost identical oscillators decreases when an impurity (parameter mismatch)
is introduced in the relay unit \cite{rb2012}.
 Further,
synchronization condition as a function of Lyapunov exponents and parameters has been obtained~\cite{aslibs2007}. 
ZLS has attracted a plethora of research activities, mainly because of its potential applications in secure
communication over a public channel \cite{ek2006}, and it was recently shown that it is possible
to use ZLS phenomenon in two mutually coupled chaotic systems for bidirectional secure communication
(where both systems sending/receiving messages simultaneously at the same time) \cite{ik2008}.
The above works clearly demonstrate the occurrence of ZLS in low-dimensional systems. 

On the other hand, synchronization in coupled time-delay systems is an intriguing phenomenon because of the 
infinite-dimensional nature of the underlying systems which exhibit hyperchaotic attractors characterized by 
multiple positive Lyapunov Exponents (LEs) even for small values of time-delay. 
Synchronizing such time-delay systems is very challenging and has 
potential applications in diverse areas involving physical, chemical, biological,
neurological and electrical systems \cite{ml2011, jnc2007, fma2010, whk2004, cl2004}.
Different types of synchronization have been recently observed 
numerically along with experimental evidence in coupled time-delay systems \cite{dvs2006, dvs2007, dvs2010, ks2011, ak1998}.

Motivated by the above works and ideas, in this paper, we report an
experimental confirmation of ZLS via dynamical relaying in coupled nonlinear time-delayed electronic 
circuits in the hyperchaotic regime corroborated by numerical simulations. 
For this purpose, we have taken three identical time-delayed electronic circuits (exhibiting hyperchaotic attractors)
and couple them with two different possible coupling configurations, namely 
mutual and subsystem coupling configurations.
Similar coupling configurations have been previously employed 
in low-dimensional chaotic systems to achieve ZLS \cite{m2010}.
In these two coupling configurations, we find that when there is a 
feedback between the relay (central) system and at least one of 
the outer systems, ZLS takes place between the two outer systems,  while the relay and outer
systems exhibit inverse phase synchronization (IPS). Snapshots of the time evolution of the 
systems and phase projection plots as observed from the oscilloscope confirm the occurrence of ZLS in
coupled time-delayed electronic circuits. Also the quality of synchronization
can be numerically verified using the correlation coefficient and the similarity function. 
It is to be noted that in the case of mutual coupling configuration ZLS between the two outer systems
occur both in the periodic and hyperchaotic regimes, whereas in the subsystem coupling configuration
it occurs only in the hyperchaotic regime. These facts are characterized and confirmed from 
the changes in the largest LEs of the coupled systems and the correlation coefficient
as a function of the coupling strength. We have also derived a sufficient stability condition
for the occurrence of ZLS using the Krasovskii-Lyapunov functional theory. Further, IPS is experimentally
confirmed using time series plots and phase coherence is characterized both 
qualitatively and quantitatively 
by the concept of localized sets and the Correlation of Probability of Recurrence (CPR), respectively.

The paper is organized as follows: In Sec.~\ref{system}, we describe the 
system employed to demonstrate the occurrence of ZLS and explain the circuit 
configuration. In Sec.~\ref{sec:level2}, we explain the existence of ZLS and characterize the 
results in mutual coupling configuration using experimental and numerical
evidences along with linear stability analysis. The occurrence of ZLS in subsystem coupling configuration
is discussed in Sec.~\ref{sec:level3} and finally Sec.~\ref{sec:level5}
is dedicated to discussion and conclusion.

\begin{figure}
\centering
\includegraphics[width=0.9\columnwidth]{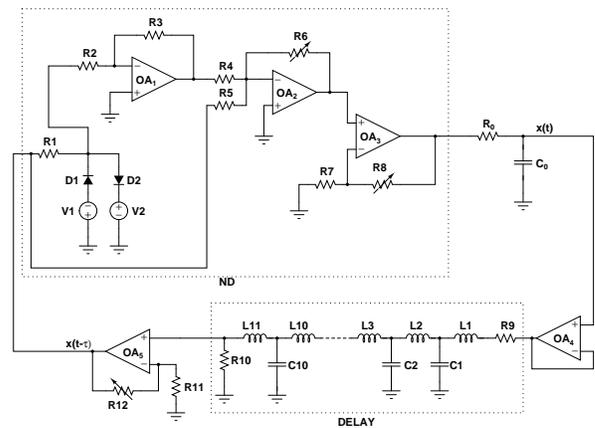}
\caption{\label{cir_dig} Circuit diagram of a single time delayed feedback oscillator with a 
nonlinear device (ND) unit, a time-delay unit (DELAY) and a low pass first-order $R_{0}C_{0}$ filter.}
\end{figure}

\section{\label{system}System description}
In this section, we briefly describe the system which we are using in this paper to demonstrate
the synchronization phenomenon. We consider the following scalar delay differential
equation
\begin{equation}
\dot{x} = -\alpha x(t)+\beta f(x(t-\tau))
\label{eqn0}
\end{equation} 
where $\alpha$, $\beta$ are the system parameters, and $\tau$ is the time-delay.
The nonlinear function $f(x)$ can be effectively implemented by a piecewise linear
function defined as 
\begin{equation}
f(x) = AF^{*}-Bx.
\label{eqn2}
\end{equation}
Here
\begin{eqnarray}
F^{*}=
\left\{
\begin{array}{cc}
-x^{*},&  x < -x^{*}  \\
            x,&  -x^{*} \leq x \leq x^{*} \\
            x^{*},&  x > x^{*}. \\ 
         \end{array} \right.
\label{eqn3}
\end{eqnarray}
The system parameters are chosen as $\alpha=1.0$, $\beta=1.2$, $\tau=6.0$, $A=5.2$, 
$B=3.5$ and $x^{*}$ is the threshold value fixed at $x^{*}=0.7$. These parameter values are used 
throughout the paper for numerical simulation. It may be noted that for the above values, 
the single system (\ref{eqn0}) exhibits a hyperchaotic attractor with multiple (four) positive 
LEs  (see Fig. 6  in  refs. \cite{ks2011, ksir2011}). 

\subsection{\label{circuit}Circuit realization}
The electronic circuit investigated here is given in Fig.~\ref{cir_dig} which describes the 
dynamics of Eq.~(\ref{eqn0}) along with the threshold nonlinear function $f(x)$ 
given by Eq.~(\ref{eqn2}). This circuit 
has a ring structure and consists of a diode based
nonlinear device (ND) unit with two amplification stages (OA$_{2}$ and OA$_{3}$), a time-delay
unit (DELAY), and a low pass first order $R_{0}C_{0}$ filter.
In this circuit, $\mu A741$s are engaged as operational amplifiers. The constant voltage sources are
$V_{1}$ and $V_{2}$, and the voltage supply for all active devices is $\pm 12 V$. 
One can adjust the threshold value of the three segment piecewise function (Eq.~(\ref{eqn3})) by altering
the voltage values $V_{1}$ and $V_{2}$.
By applying Kirchhoff's laws to this circuit (Fig.~\ref{cir_dig}), the state equation 
can be written as $R_{0}C_{0}\frac{dU(t)}{dt}=-U(t)+F[k_{f}(U(t-T_{d}))]$, where $U(t)$
is the voltage across the capacitor $C_{0}$, $U(t-T_{d})$ is the voltage across
the delay unit, $T_{d}(=n\sqrt{LC})$ is the delay time, $n$ is the number of 
LC filter units, and $F[k_{f}(U(t-T_{d}))]$ is the static characteristic of the ND unit.
\begin{figure}
\centering
\includegraphics[width=0.8\columnwidth]{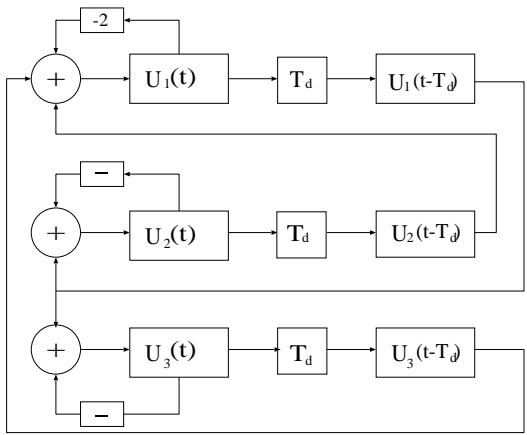}
\caption{\label{mut_cup} Circuit block diagram of the three coupled time delayed feedback 
oscillator for mutual coupling configuration (\ref{mut_eqn}).}
\end{figure}

To analyze the above circuit equation, we transform it to the dimensionless oscillator
(\ref{eqn0}) by defining the dimensionless 
variables and parameters as $x(t)=\frac{U(t)}{U_{s}}$, $t^{\prime}=\frac{t}{R_{0}C_{0}}$,
$\tau=\frac{T_{d}}{R_{0}C_{0}}$, $\alpha=1.0$, $k_{f}=\beta$, and $t^{\prime}\rightarrow t$. 
The circuit parameters are fixed as follows:
$R_{1}=1K\Omega$, $R_{2}=R_{3}=10K\Omega$, $R_{4}=2K\Omega$, $R_{5}=3K\Omega$, $R_{6}=10.4K\Omega$ (trimmer-pot),
$R_{7}=1K\Omega$, $R_{8}=5K\Omega$ (trimmer-pot), $(R_{9}=R_{10}=1K\Omega$, $R_{11}=10K\Omega$,
$R_{12}=20K\Omega$ (trimmer-pot), $R_{0}=1.86K\Omega$, $C_{0}=100nF$, $L_{i}=12mH$ ($i=1,2,\cdots,11$),
$C_{i}=470nF$ ($i=1,2,\cdots,10$), $n=10$. $T_{d}=0.751ms$, $R_{0}C_{0}=0.268ms$, and so the 
time-delay $\tau\approx2.8$ for the chosen values of the circuit parameters.

The bifurcation scenario and the dynamics of this time-delayed chaotic oscillator 
has been investigated in some detail in Ref.~\cite{ksir2011}. In the following 
sections, we will demonstrate the existence
of zero lag synchronization in coupled time-delayed piecewise linear 
electronic circuits of the above form for two different coupling configurations.

\section{\label{sec:level2}Mutual coupling configuration}
In this section, we demonstrate the occurrence of ZLS in coupled time-delay systems
with mutual coupling configuration. For this purpose, we have considered three identical
time-delayed electronic circuits each of the form of Fig.~\ref{cir_dig} and 
coupled them as shown in the block diagram of Fig.~\ref{mut_cup}. 
The state equations for the coupled electronic circuit (Fig.~\ref{mut_cup}) can be written as
\begin{subequations}
\begin{eqnarray}
R_{0}C_{0} \frac{dU_{1}(t)}{dt} &= -U_{1}(t)+f[k_{f}U_{1}(t-T_{d})]+ \nonumber \\
& \varepsilon^{\prime}[U_{2}(t-T_{d})-2U_{1}(t)+U_{3}(t-T_{d})], \\
R_{0}C_{0} \frac{dU_{2}(t)}{dt} &= -U_{2}(t)+f[k_{f}U_{2}(t-T_{d})]+ \nonumber \\
& \varepsilon^{\prime}[U_{1}(t-T_{d})-U_{2}(t)], \\
R_{0}C_{0} \frac{dU_{3}(t)}{dt} &= -U_{3}(t)+f[k_{f}U_{3}(t-T_{d})]+ \nonumber \\
& \varepsilon^{\prime}[U_{1}(t-T_{d})-U_{3}(t)],
\end{eqnarray}
\label{mut_eqn}
\end{subequations}
where the variables $U_{1}(t)$, $U_{2}(t)$ and $U_{3}(t)$ correspond to the output
variables of each circuit. By defining the normalized variables 
and parameters as given in Sec.~\ref{circuit} and $\varepsilon^{\prime}=\varepsilon$, 
one obtains the equivalent dimensionless equation as follows:
\begin{subequations}
\begin{eqnarray}
\dot{x}_1(t) &=& -\alpha x_{1}(t)+\beta f(x_{1}(t-\tau))+ \nonumber \\
& & \varepsilon[x_{2}(t-\tau)-2x_{1}(t)+x_{3}(t-\tau)],\\
\dot{x}_2(t) &=& -\alpha x_{2}(t)+\beta f(x_{2}(t-\tau))+ \nonumber \\
& & \varepsilon[x_{1}(t-\tau)-x_{2}(t)],\\
\dot{x}_3(t) &=& -\alpha x_{3}(t)+\beta f(x_{3}(t-\tau))+\nonumber \\
& & \varepsilon[x_{1}(t-\tau)-x_{3}(t)].
\end{eqnarray}
\label{eqn1}
\end{subequations}
Here, $x_{1}(t)$ is the central relay system, $x_{2}(t)$ and $x_{3}(t)$ are the two
outer systems and $\varepsilon$ is the coupling strength between the systems.
\begin{figure}
\centering
\includegraphics[width=0.7\columnwidth]{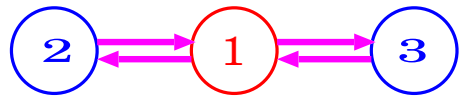}
\caption{\label{fig2} (Color online) Schematic diagram for mutual coupling configuration.}
\end{figure}

In this coupling configuration, the relay unit (system 1) sends its delayed 
signal to the outer units (systems 2 and 3)
and both the outer units also send their delayed feedback to the relay unit (Fig.~\ref{mut_cup}).
For simple illustration, the schematic diagram for this coupling configuration is shown in Fig.~\ref{fig2}. 
\subsection{\label{numerical}Experimental and numerical results}

In the absence and for lower values of coupling strength,
the three systems evolve freely according to their own dynamics. For sufficiently
large values of coupling strength, if the two outer systems are directly connected
without the relay unit (system 1), they get synchronized with a lead or lag time
equal to the coupling delay \cite{aw2007, jm2004, lbs2006}. On the other hand, if we couple them through the relay
system (as depicted in Fig.~\ref{fig2}), then remarkably both the outer systems are synchronized without
a time lag (zero-lag). In addition, the outer systems (2 and 3) exhibit IPS with the relay system. 
\begin{figure}
\centering
\includegraphics[width=0.8\columnwidth]{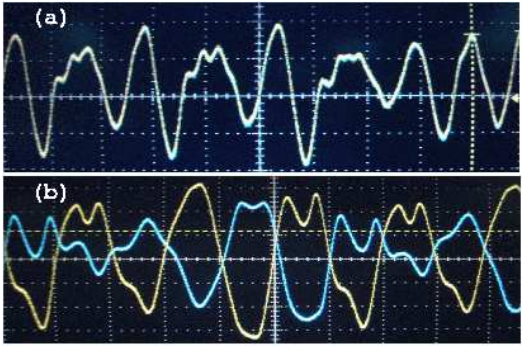}
\caption{\label{fig4} (Color online)Experimental verification of ZLS and IPS 
in mutual coupling configuration. (a) Time evolution of both the outer circuits (2 and 
3) displays ZLS ($U_{2}(t)$ - yellow curve and $U_{3}(t)$ - blue curve); vertical scale 1 unit = $2 V$,
horizontal scale 1 unit = $1 ms$ and (b) time evolution of the relay and one of the outer circuits shows IPS
($U_{1}(t)$ - yellow curve and $U_{2}(t)$ - blue curve); vertical scale 1 unit = $2 V$, horizontal scale 1 unit = $2 ms$.}
\end{figure}
\begin{figure}
\centering
\includegraphics[width=0.9\columnwidth]{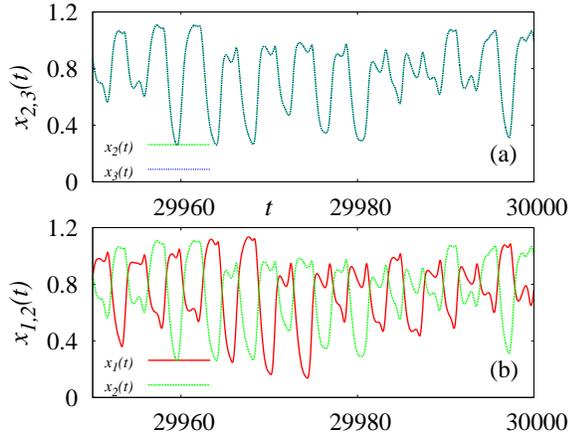}
\caption{\label{fig3} (Color online) Numerically obtained time series of the systems 
in mutual coupling configuration (Eq.~(\ref{eqn1}))
for $\varepsilon=1.55$. (a) The two outer systems $x_{2}(t)$ and $x_{3}(t)$ 
display ZLS, (b) the outer system $x_{2}(t)$ shows IPS with the relay system $x_{1}(t)$.}
\end{figure}

\begin{figure}
\centering
\includegraphics[width=0.9\columnwidth]{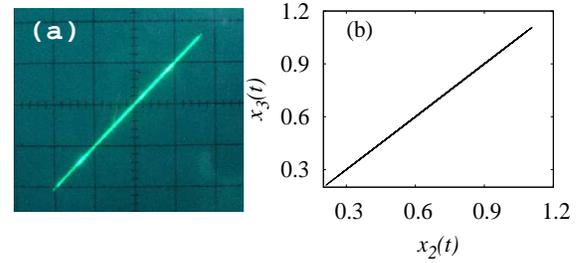}
\caption{\label{fig5} (Color online) (a)  Experimental (x-axis: voltage $U_{2}(t)$ (1 unit = 2.0 V); 
y-axis: $U_{3}(t)$ (1 unit = 2.0 V)) and (b) numerical realization of the phase
portraits of the systems $x_{2}(t)$ and $x_{3}(t)$ showing ZLS.}
\end{figure}

Snapshots of wave forms of the circuits as seen in the oscilloscope are shown in Fig.~\ref{fig4}.
ZLS between the two outer circuits is presented in 
Fig.~\ref{fig4}(a) and the realization of IPS between one of the outer and the central circuits 
is given in Fig.~\ref{fig4}(b).
Figure~\ref{fig3} shows the numerical simulation of Eq.~(\ref{eqn1}) 
for the coupling strength $\varepsilon=1.55$. 
Systems 2 and 3 maintain a perfect ZLS as depicted in Fig.~\ref{fig3}(a). 
ZLS is further confirmed from the phase 
portraits of the corresponding systems. Figures~\ref{fig5}(a) and \ref{fig5}(b) 
show experimental and numerical 
realizations of the phase portraits of the two outer systems, respectively.
The diagonal line represents the existence of complete (zero-lag) 
synchronization between the two outer systems 2 and 3.
Further, the outer system 2 is in IPS with the relay system 1, 
where essentially the extrema of the time series of the two systems occur opposite to each other as 
clearly displayed in Figs.~\ref{fig4}(b) and \ref{fig3}(b).
\begin{figure}
\centering
\includegraphics[width=0.8\columnwidth]{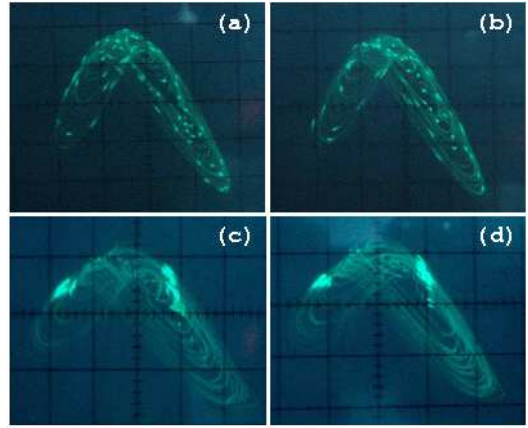}
\caption{\label{fig5a} (Color online) Experimental realization of the framework 
of localized sets for mutual coupling configuration. 
(a), (b) The sets are spread over the attractors indicating the absence of 
phase coherence for lower values of coupling strength. (c), (d) For a sufficiently large 
value of coupling strength, the sets are localized on the attractors which
indicates phase coherence. In (a), (c) x-axis: voltage $U_{1}(t)$ (1 unit = 0.5 V), 
y-axis: voltage $U_{1}(t-T_{d})$ (1 unit = 2.0 V) and in (b), (d) x-axis: voltage $U_{2}(t)$ (1 unit = 0.5 V), 
y-axis: voltage $U_{2}(t-T_{d})$ (1 unit = 2.0 V).}
\end{figure}
\begin{figure}
\centering
\includegraphics[width=1.0\columnwidth]{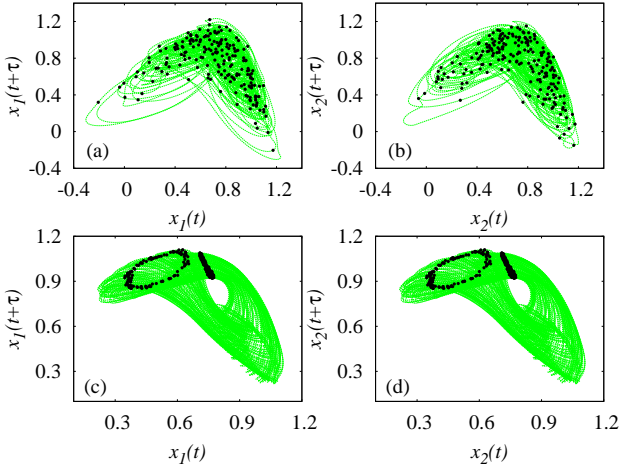}
\caption{\label{fig5b} (Color online) Numerically obtained equivalent figures for Fig.~\ref{fig5a}. 
(a), (b) correspond to the absence of phase coherence for $\varepsilon=0.5$. In
(c), (d) the sets are localized on the attractors indicating 
phase locking for $\varepsilon=1.55$.}
\end{figure}
\begin{figure}
\centering
\includegraphics[width=1.0\columnwidth]{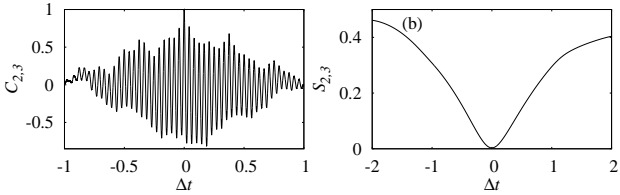}
\caption{\label{fig6} ZLS of the two outer systems 2 and 3 are confirmed by 
(a) the correlation coefficient (Eq.~\ref{eqn4a}) and (b) the similarity function (Eq.~\ref{eqn4b}) for $\varepsilon=1.55$.}
\end{figure}
\begin{figure}
\centering
\includegraphics[width=0.95\columnwidth]{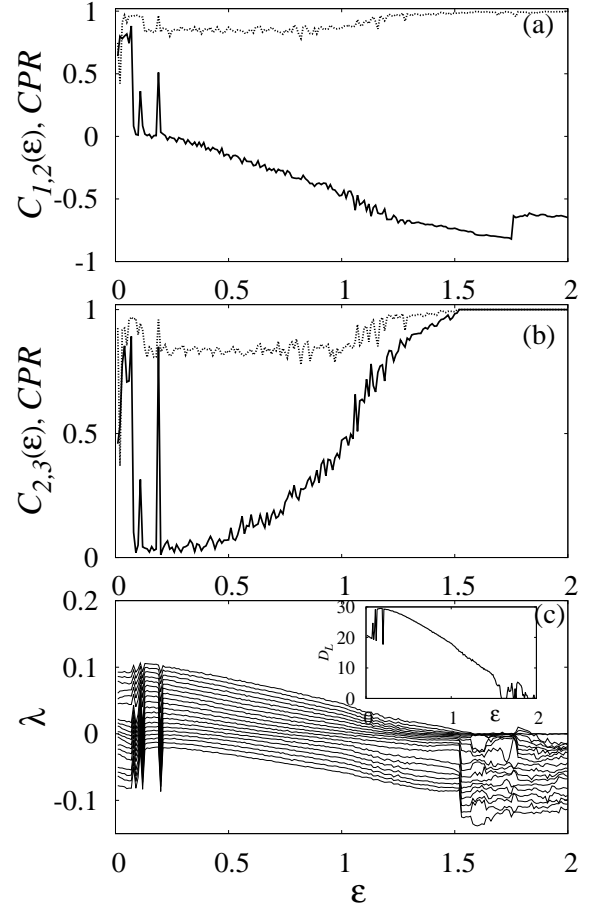}
\caption{\label{fig7} (a) Correlation coefficient $C_{1,2}(\varepsilon)$ (continuous curve) and the
index CPR (dotted curve) of the systems 1, 2, (b) correlation coefficient $C_{2,3}(\varepsilon)$
and the CPR of the two outer systems 2, 3 and (c) display the transition in the 
largest LEs of the coupled systems as a function of the coupling strength in mutual coupling configuration.
The inset figure in (c) shows the Kaplan-Yorke dimension ($D_{L}$) as a function of the
coupling strength.}
\end{figure}

A qualitative measurement of phase coherence can be visualized 
both experimentally and numerically by using the framework of localized sets \cite{tpmsb2007}.
The basic idea of this characterization is that the set of points obtained by sampling the time series
of the system 2 whenever a maximum occurs in the relay system is plotted along
with the attractor of the system 2 and vice-versa. Depending upon the property of the set,
one can find whether phase coherence exists or not. The coupled systems are said to be phase synchronized
upon localization of the observed sets on the attractor. On the other hand, the sets that 
spread over the entire attractor confirm asynchronization. This approach provides a general 
way to identify phase synchronization even in non-phase-coherent attractors. 

For the experimental implementation of the framework of localized sets, 
maxima of the outer system  and minima of the
relay units are taken as reference. Using the circuit given 
in Fig.~7.12 in  pp. 147 of Ref.~\cite{mlkm1995},
we generate the impulse whenever the
input signal of the outer (relay) system attains maximum (minima). While the
attractor of the relay (outer) system is in the X-Y channel of the
oscilloscope, we feed the impulse signal to Z- input. Whenever the
impulse hits the attractor of the relay (outer) unit one can see the bright spot on the
attractor of the relay (outer) system. 
As pointed out above, if the both systems are in CPS the spots (sets) are
localized on the attractors otherwise the sets are spread  over the
entire attractor.

Figures \ref{fig5a} (a) and \ref{fig5a} (b) show the experimentally obtained attractors along with the sets
of the relay system and the system 2, respectively. Here the sets are distributed over the 
entire attractor for a low value of coupling strength related to the absence of phase synchronization.
The corresponding numerically obtained figures are plotted in Figs.~\ref{fig5b} (a) 
and \ref{fig5b} (b) for the value of coupling strength $\varepsilon=0.5$.
If we increase the coupling strength to sufficiently large value,
the sets are localized on the attractor as depicted in Figs.~\ref{fig5a} (c) and \ref{fig5a} (d) 
confirming the phase locking of both systems. The corresponding numerical figures 
are plotted in Figs.~\ref{fig5b} (c) and \ref{fig5b} (d) for the value of $\varepsilon=1.55$.
\begin{figure}
\centering
\includegraphics[width=0.8\columnwidth]{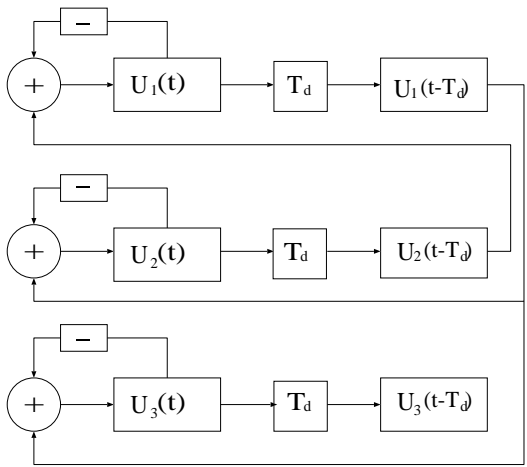}
\caption{\label{sub_cup} Circuit block diagram of the three coupled time delayed feedback oscillator
for the subsystem coupling configuration (\ref{sub_eqn}).}
\end{figure}

In order to characterize the quality of synchronization and to find out the lag between the 
time series of the two outer systems, we study the correlation coefficient
and the similarity function between the outer systems. The correlation coefficient 
and the similarity function, respectively, are given by the expressions
%
\begin{eqnarray}
C_{2,3} &=& \frac{\langle(x_{2}(t)-\langle x_{2}(t)\rangle)(x_{3}(t+\Delta t)-\langle x_{3}(t)\rangle)\rangle}{\sqrt{\langle(x_{2}(t)-\langle x_{2}(t)\rangle)^{2}\rangle \langle(x_{3}(t)-\langle x_{3}(t)\rangle)^{2}\rangle}},\label{eqn4a}\\
S_{2,3} &=& \frac{\langle(x_{3}(t+\tau)-x_{2}(t))^{2}\rangle}{\sqrt{\langle x_{2}^{2}(t)\rangle \langle x_{3}^{2}(t)\rangle}},
\label{eqn4b}
\end{eqnarray}
%
where the $\langle \quad \rangle$ brackets indicate time averaging. Using Eq.~(\ref{eqn4a}), we compute the correlation between the
outer systems as a function of time shifts ($\Delta t$), characterizing the quality of the synchronization. The time lag
between the two outer systems can be determined by the position of the global maximum of the correlation coefficient for ZLS. 
The global maximum of the correlation coefficient has a value close to unity, 
at $\Delta t=0$, which reflects that there is no delay between both the outer systems. 
In the case of the similarity function (Eq.~(\ref{eqn4b})), if $x_{2}(t)=x_{3}(t)$ (for ZLS), 
then the similarity function reaches the minimum $S\approx0$ for time shift $\tau=0$. 
For a nonzero value of the time shift $\tau$, $S\approx0$ corresponds to a lag between the 
two signals $x_{2}(t)$ and $x_{3}(t)$. Figure~\ref{fig6} (a) shows the correlation coefficient of the systems 2 and 3 as a 
function of $\Delta t$ with its global maximum ($C_{2,3}\approx1.0$) at $\Delta t=0$
confirming the ZLS between them for the coupling strength $\varepsilon=1.55$. Similarly, 
the similarity function
of the two outer systems 2 and 3 is depicted in Fig.~\ref{fig6} (b) which clearly indicates 
that the minimum $S_{2,3}\approx0$ occurs at
$\tau=0$ again confirming the existence of ZLS between the systems $x_{2}(t)$ and $x_{3}(t)$ .

The transition from asynchronization to ZLS between the outer systems is 
further characterized by calculating the correlation coefficient 
(with $\Delta t=0$ in Eq.~(\ref{eqn4a})) and the transition in the largest LEs 
of the coupled time-delay systems as a function of the coupling
strength in the range $\varepsilon \in (0,2)$. Further, phase coherence is 
characterized by using one of the recurrence quantifications, namely 
the Correlation of Probability of Recurrence (CPR) \cite{mcrmt2005, nmmcr2007}.
If the phases of the coupled systems are entrained (mutually locked) completely,
then the probability of the recurrence is
maximal at a time and CPR $\approx 1$. In contrast, one can expect a drift in
the probability of recurrences resulting in low values of  CPR 
characterizing the degree of locking between the coupled systems~\cite{mcrmt2005, nmmcr2007}.
Figures~\ref{fig7}(a) and \ref{fig7}(b) display the correlation coefficient 
(continuous curve) and the index CPR (dotted curve) between the systems 1, 2 
and systems 2, 3, respectively. This way, we can distinguish the following regimes:

(i) For certain low values of $\varepsilon$, all three systems exhibit in-phase
synchronization where CPR $\approx 1$. Further, the high degree of the
correlation coefficients $C_{1,2}(\varepsilon)$ and $C_{2,3}(\varepsilon)$ 
but with values less than unity quantifies the degree of correlation in
their amplitudes, corroborating the existence of approximate complete synchronization
between all the three systems. This synchronization transition is also confirmed
by the changes in the largest LEs of the coupled systems (\ref{eqn1}). 
In Fig.~\ref{fig7}(c), we have plotted the $19$ largest LEs 
as a function of the coupling strength $\varepsilon$. For $\varepsilon=0$, 
there are $12$ positive LEs (each system has four positive LEs). We
note that for nonzero low values of $\varepsilon$ there exists 
only $9$ positive LEs confirming the transition to the approximate 
synchronization with three of the positive LEs becoming negative in the 
above range of $\varepsilon$.

(ii) In the intermediate range of $\varepsilon$, $C_{1,2}$ decreases and $C_{2,3}$
increases upon increasing $\varepsilon$ indicating the loss of correlation between the relay
and outer units, while the index CPR oscillates around the value $0.8$ among all the
three systems quantifying the degree of drift in their phases.
On the other hand, the degree of correlation between the outer units  
increases at the same time leading to the transition to ZLS from 
approximate complete synchronization. Further,
the largest Lyapunov exponents gradually acquire values less than zero, which
also confirms the above synchronization transition.

(iii) If we increase the coupling strength further, the index CPR of the systems 1, 2 
increases and reaches almost unity for $\varepsilon>1.4$, whereas their 
correlation coefficient $C_{1,2}(\varepsilon)$ becomes negative, which indeed
confirms the inverse phase synchronization between units 1 and 2 [Fig.~\ref{fig7}(a)]. 
On the other hand, the correlation coefficient $C_{2,3}(\varepsilon)$ and 
the index CPR of the two outer systems increase as a function of $\varepsilon$ 
and for $\varepsilon=1.51$ both 
reach the unit value corroborating the occurrence of ZLS between them [Fig.~\ref{fig7}(b)].

(iv)  Further, except for the largest three positive LEs, all the other LEs of the coupled system 
become negative for $\varepsilon > 1.51$
confirming the existence of ZLS between 
the outer units in the chaotic regime, which is the common synchronization 
manifold  arising due to mutual interaction among all the systems. We also note that
there exist windows of ZLS in periodic regimes,  where all the LEs are less than zero.
It is worth to emphasize at this point that we have observed ZLS where
the synchronization manifold is both periodic and chaotic depending on
the value of the coupling strength $\varepsilon$ as unveiled  by the Lyapunov exponents.
Further, the Kaplan-Yorke dimension ($D_{L}$) \cite{jdf1982, jk1979} is depicted 
in the inset in Fig.~\ref{fig7}(c) (defined as the sum of all the positive LEs)
as a function of the coupling strength, which clearly indicates the 
periodic ($D_{L}=0$) and chaotic regimes ($D_{L}>0$).

\subsection{\label{stability}Linear stability analysis for ZLS}
In this subsection, we analytically investigate the existence of ZLS between the 
two outer oscillators in the mutual coupling configuration [Eq.~(\ref{eqn1})(b, c)]. For this, we
consider the time evolution of the difference system with the state variable $\Delta=x_{2}(t)-x_{3}(t)$
(which corresponds to the ZLS manifold of the two outer systems) for small values
of $\Delta$. The evolution equation for the synchronization manifold can be written as
\begin{equation}
\dot{\Delta}=-(\alpha+\varepsilon)\Delta+\beta [A F^{\ast\prime}(x_{2}(t-\tau))-B] \Delta_{\tau}, \quad \Delta_{\tau}=\Delta(t-\tau)
\label{eqn1a}
\end{equation}
so that the stability condition can be deduced analytically. The synchronization manifold  
is locally attracting if the origin of (\ref{eqn1a}) is stable. 
From the Krasovskii-Lyapunov theory \cite{kkk1963},
one can define a continuous positive definite Lyapunov functional of the form
\begin{equation}
V(t)=\frac{1}{2}\Delta^{2}+\mu \int_{-\tau}^{0} \Delta^{2}(t+\theta)d\theta, \qquad V(0)=0,
\label{eqn1b}
\end{equation} 
where $\mu$ is an arbitrary positive parameter $\mu>0$. The above Lyapunov function $V(t)$
approaches zero as $\Delta \rightarrow 0$. The derivative of the above equation (\ref{eqn1b})
along the trajectory of the synchronization manifold (\ref{eqn1a}),
\begin{equation}
\frac{dV}{dt}=-[\alpha+\varepsilon] \Delta^{2}+\beta [AF^{\ast\prime}(x_{2}(t-\tau))-B]\Delta \Delta_{\tau}+\mu\Delta^{2}-\mu\Delta_{\tau}^{2},
\label{eqn1c}
\end{equation}
should be negative for the stability of the synchronization manifold $\Delta=0$. 
This requirement results in the condition for stability as
\begin{equation}
(\alpha+\varepsilon)>\frac{\beta^{2}}{4\mu}[AF^{\ast\prime}(x_{2}(t-\tau))-B]^{2}+\mu=\Phi(\mu).
\label{eqn1d}
\end{equation}
One may note that $\Phi(\mu)$ as a function of $\mu$ for a given $F^{\ast\prime}(x)$
 has an absolute minimum at $\mu={|\beta [A F^{\ast\prime}(x_{2}(t-\tau))-B]|}/2$ 
with $\Phi_{min}=|\beta [A F^{\ast\prime}(x_{2}(t-\tau))-B]|$.
Since $\Phi \geq \Phi_{min}=|\beta [A F^{\ast\prime}(x_{2}(t-\tau))-B]|$, from the inequality (\ref{eqn1d}),
a {\it sufficient} condition for the asymptotic stability is 
\begin{equation}
(\alpha+\varepsilon)>|\beta (A F^{\ast\prime}(x_{2}(t-\tau))-B)|.
\label{eqn1e}
\end{equation}
Now, from the form of the piecewise linear function $f(x)$ in Eq.~(\ref{eqn3}), we have,
\begin{eqnarray}
|F^{\ast\prime}(x_{2}(t-\tau))|=
\left\{
\begin{array}{cc}
            0,&  |x| > x^{*}  \\
            1,&  |x| \leq x^{*}. \\
         \end{array} \right.
\label{eqn1f}
\end{eqnarray}
Consequently, the stability condition (\ref{eqn1e}) becomes 
$(\alpha+\varepsilon)>|\beta (A-B)|$, corresponding to the inner 
regime $|x| \leq x^{*}$ where most of the dynamics is confined, 
for the asymptotic synchronized state $\Delta=0$. 
From our extensive numerical analysis as noted above, we find that ZLS between the two outer
oscillators occur for the coupling strength $\varepsilon=1.51$ satisfying the stability condition
$\varepsilon>|\beta (A-B)|-\alpha=1.04)$. However, we also note that
the stability condition  corresponding to the outer regime $|x| > x^{*}$, that is,
$(\alpha+\varepsilon)>|\beta B|$ is not validated by our numerical results 
in achieveing ZLS, essentially because the stability condition obtained
from the Lyapunov functional theory is only a sufficiency condition. 
Moreover, the role of the relay unit is not captured by the evolution equation (\ref{eqn1a}) 
for the ZLS manifold in the above stability analysis. 

\section{\label{sec:level3}Subsystem coupling configuration}
\begin{figure}
\centering
\includegraphics[width=0.7\columnwidth]{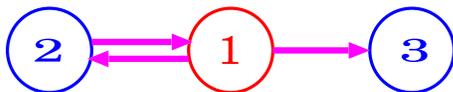}
\caption{\label{fig8} (Color online) Schematic diagram for the subsystem coupling configuration.}
\end{figure}

Next we consider the coupling configuration, where the relay system sends its delayed signal to the two outer
systems and only one system (here system 2) sends its delayed feedback to the relay system. This configuration is called
a subsystem coupling. The circuit for the subsystem coupling
configuration is shown in Fig.~\ref{sub_cup} as a block diagram. The state equations for the 
coupled electronic circuit (Fig.~\ref{sub_cup}) can be written as follows:
\begin{subequations}
\begin{eqnarray}
R_{0}C_{0} \frac{dU_{1}(t)}{dt} &= -U_{1}(t)+f[k_{f}U_{1}(t-T_{d})]+ \nonumber \\
& \varepsilon^{\prime}[U_{2}(t-T_{d})-U_{1}(t)], \\
R_{0}C_{0} \frac{dU_{2}(t)}{dt} &= -U_{2}(t)+f[k_{f}U_{2}(t-T_{d})]+ \nonumber \\
& \varepsilon^{\prime}[U_{1}(t-T_{d})-U_{2}(t)], \\
R_{0}C_{0} \frac{dU_{3}(t)}{dt} &= -U_{3}(t)+f[k_{f}U_{3}(t-T_{d})]+ \nonumber \\
& \varepsilon^{\prime}[U_{1}(t-T_{d})-U_{3}(t)].
\end{eqnarray}
\label{sub_eqn}
\end{subequations}
The equivalent dimensionless equations of motion (see Sec.~\ref{system}) 
for the above configuration can be given as
\begin{subequations}
\begin{eqnarray}
\dot{x}_1(t) &= -\alpha x_{1}(t)+\beta f(x_{1}(t-\tau))+ \nonumber \\ 
& \varepsilon[x_{2}(t-\tau)-x_{1}(t)],\\
\dot{x}_2(t) &= -\alpha x_{2}(t)+\beta f(x_{2}(t-\tau))+ \nonumber \\
& \varepsilon[x_{1}(t-\tau)-x_{2}(t)],\\
\dot{x}_3(t) &= -\alpha x_{3}(t)+\beta f(x_{3}(t-\tau))+ \nonumber \\
& \varepsilon[x_{1}(t-\tau)-x_{3}(t)].
\end{eqnarray}
\label{eqn5}
\end{subequations}
The schematic diagram for this configuration is sketched in Fig.~\ref{fig8}. 
In the absence of the coupling ($\varepsilon=0$), all the three systems evolve 
independently and  therefore there is no synchronization between them. 
As the coupling strength increases ($\varepsilon>0$) the two outer systems
achieve ZLS, while exhibiting IPS with the relay system.

\begin{figure}
\centering
\includegraphics[width=0.8\columnwidth]{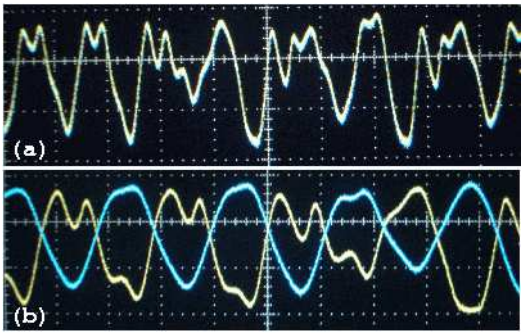}
\caption{\label{fig10} (Color online) Experimental verification of ZLS and IPS  
in subsystem coupling configuration. (a) Time evolution of both the outer circuits ($U_{2}(t)$ - yellow curve
and $U_{3}(t)$ - blue curve) displaying ZLS; x-axis: 1 unit - $2 ms$, y-axis: 1 unit - $2 V$
and (b) time series of the relay and one of the outer circuits showing IPS 
(yellow curve $U_{1}(t)$ and blue curve $U_{2}(t)$). x-axis: 1 unit - $2 ms$, y-axis: 1 unit - $2 V$.}
\end{figure}

\begin{figure}
\centering
\includegraphics[width=0.9\columnwidth]{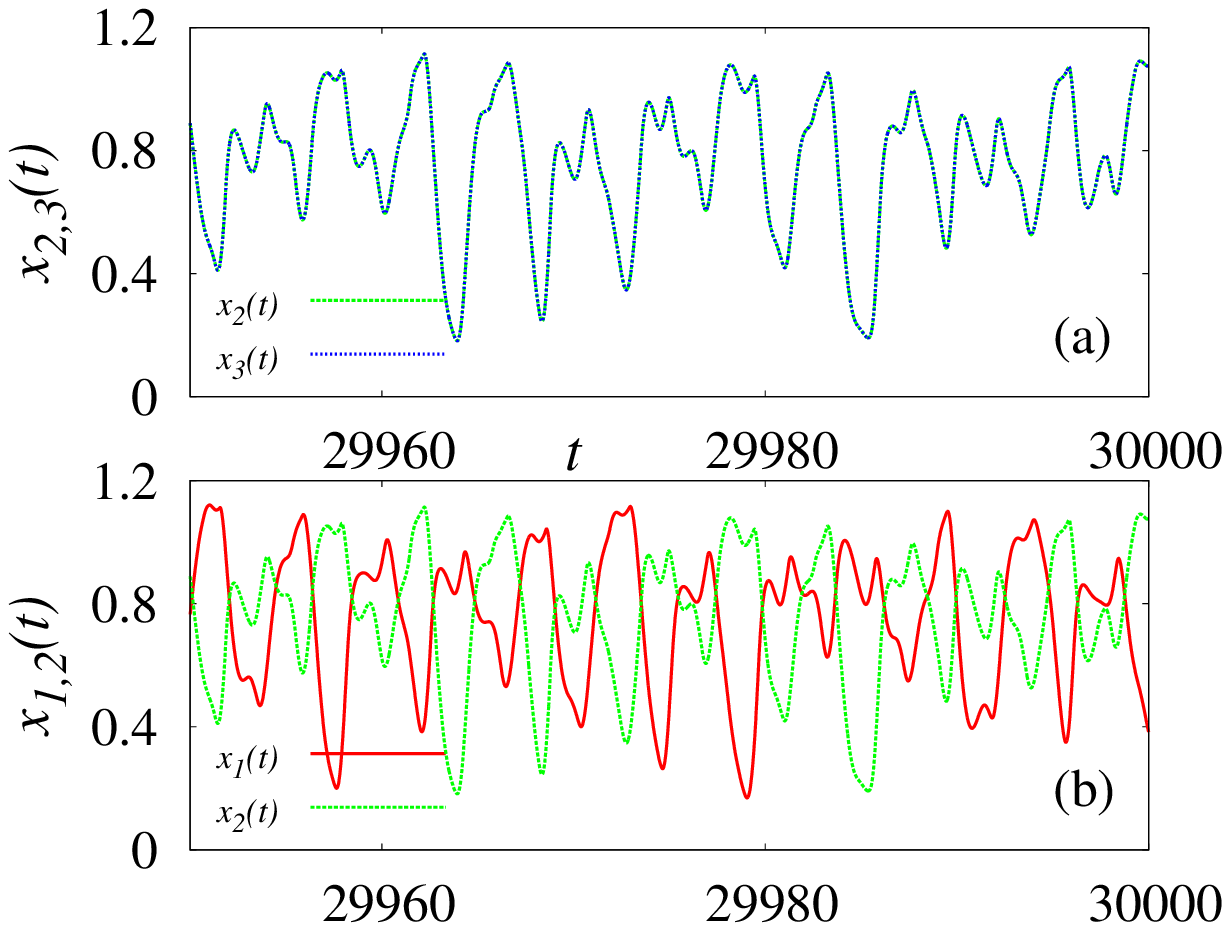}
\caption{\label{fig9} (Color online) Numerically obtained time series of the systems in 
subsystem coupling configuration (Eq.~(\ref{eqn5}))
for $\varepsilon=1.6$. (a) The two outer systems $x_{2}(t)$ and $x_{3}(t)$ 
display ZLS, while (b) the outer system $x_{2}(t)$ shows IPS with the relay system $x_{1}(t)$.}
\end{figure}

\begin{figure}
\centering
\includegraphics[width=0.9\columnwidth]{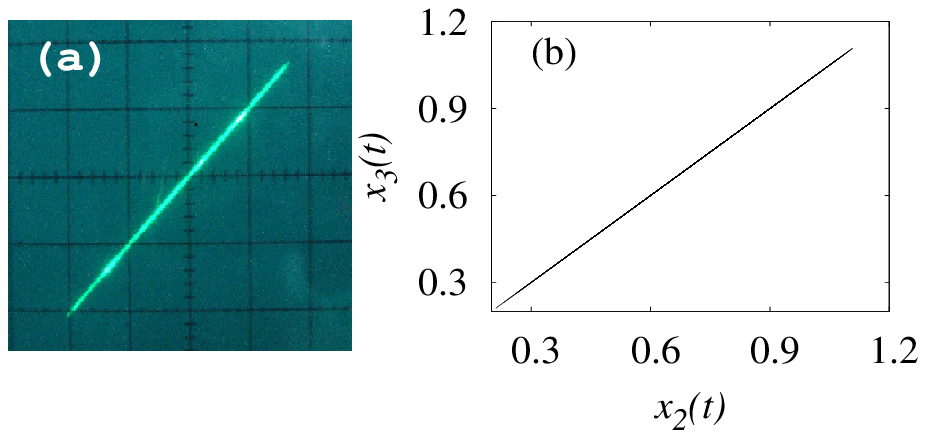}
\caption{\label{fig11} (Color online) (a) Experimental (x-axis: voltage $U_{2}(t)$ (1 unit = 2.0 V),
y-axis: voltage $U_{3}(t)$ (1 unit = 2.0 V)) and (b) numerical realization of the phase
portraits of the systems $x_{2}(t)$ and $x_{3}(t)$ in subsystem coupling configuration showing ZLS.}
\end{figure}

The experimental wave forms for suitable $\varepsilon$ is shown in Fig.~\ref{fig10}, where the time evolution of the 
two outer circuits is displayed in Fig.~\ref{fig10}(a) exhibiting ZLS. The time evolution
of the relay and one of the outer circuits (1 and 2) is displayed in Fig.~\ref{fig10}(b) confirming IPS.
The time evolution of all the three systems is also obtained using numerical simulations and depicted in Fig.~\ref{fig9}
for the value of coupling strength $\varepsilon=1.6$. Figure~\ref{fig9}(a) shows the time 
traces of the two outer systems, displaying ZLS.
The occurrence of IPS between the 
relay and outer unit (systems 1 and 2) can also be clearly seen in Fig.~\ref{fig9}(b)
where the maxima of the time series of the two systems are exactly opposite to each other. 
Figures~\ref{fig11} (a) and \ref{fig11} (b) show the phase portraits (obtained by experimental and numerical simulations, respectively) of the outer systems (2 and 3) which display ZLS. 

\begin{figure}
\centering
\includegraphics[width=0.8\columnwidth]{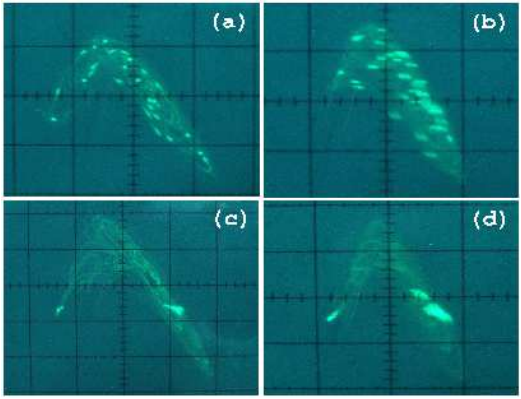}
\caption{\label{fig11a} (Color online)  Experimental realization of the
framework of localized sets for subsystem coupling configuration. (a), (b) The
sets are spread over the entire attractors indicating the absence of phase coherence
for lower values of coupling strength. (c), (d) For sufficiently large coupling 
value the sets are localized on the attractors which indicates the phase locking.
In (a), (c) x-axis: voltage $U_{1}(t)$ (1 unit = 0.5 V), y-axis: voltage $U_{1}(t-T_{d})$ (1 unit = 2.0 V)
and in (b), (d) x-axis: voltage $U_{2}(t)$ (1 unit = 0.5 V), y-axis: voltage $U_{2}(t-T_{d})$ (1 unit = 2.0 V).}
\end{figure}

\begin{figure}
\centering
\includegraphics[width=1.0\columnwidth]{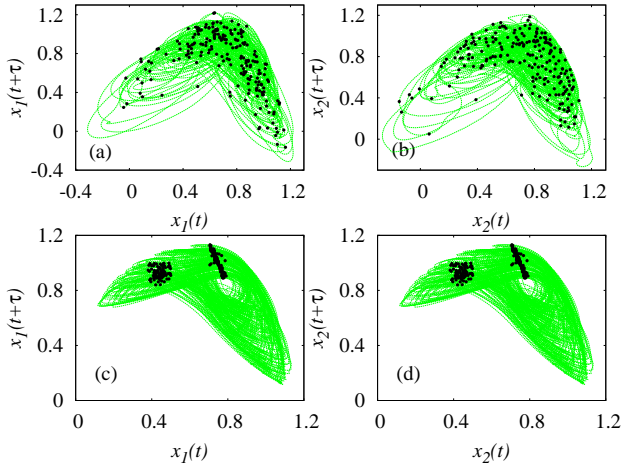}
\caption{\label{fig11b} (Color online) Numerically obtained equivalent figures for 
Fig.~\ref{fig11a}. (a), (b) correspond to the absence of phase cohernece for $\varepsilon=0.5$,
(c), (d) the sets are localized on the attractors related to the occurrence of IPS
for $\varepsilon=1.6$.}
\end{figure}

The phase coherence between the outer and the relay systems can again be visualized by plotting
the localized sets. Figure~\ref{fig11a} shows the experimentally obtained attractors along with the sets of the
relay system and the system 2. In Figs.~\ref{fig11a} (a) and \ref{fig11a} (b) the 
sets are distributed over the entire attrator for lower values of coupling strength $\varepsilon$ due to
the absence of phase synchronziation. The corresponding numerically obtained figures are plotted in 
Figs.~\ref{fig11b} (a) and \ref{fig11b} (b) for $\varepsilon=0.5$. If we increase the 
coupling strength to a sufficiently large value, the sets are localized on the
attractor as depicted in Figs.~\ref{fig11a} (c) and \ref{fig11a} (d) confirming the phase
locking of both the systems. The corresponding numerically obtained figures are plotted in 
Figs.~\ref{fig11b} (c) and \ref{fig11b} (d) for the value of $\varepsilon=1.6$.

\begin{figure}
\centering
\includegraphics[width=1.0\columnwidth]{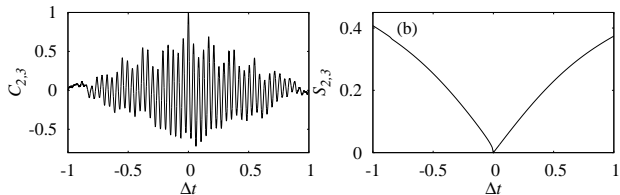}
\caption{\label{fig12} Confirmation of ZLS of the two outer systems by
(a) correlation coefficient and (b) the similarity function for $\varepsilon=1.6$ in subsystem coupling configuration.}
\end{figure}

As before, the existence of ZLS is further characterized and confirmed by calculating the 
correlation coefficient (\ref{eqn4a}) and the similarity
function (\ref{eqn4b}) of the systems 2 and 3. The global maximum of the correlation of the systems 2 and 3 
has a value close to unity ($C_{2,3}\approx0.99$) at $\Delta t=0$ indicating a complete synchronized 
behavior and that there is no lag between the two
systems (Fig.~\ref{fig12}(a)) for $\varepsilon=1.6$. Figure~\ref{fig12}(b) shows 
the minimum of the similarity function $S_{2,3}\approx0$ at $\tau=0$, 
which also indicates that both the systems exhibit ZLS.
\begin{figure}
\centering
\includegraphics[width=0.9\columnwidth]{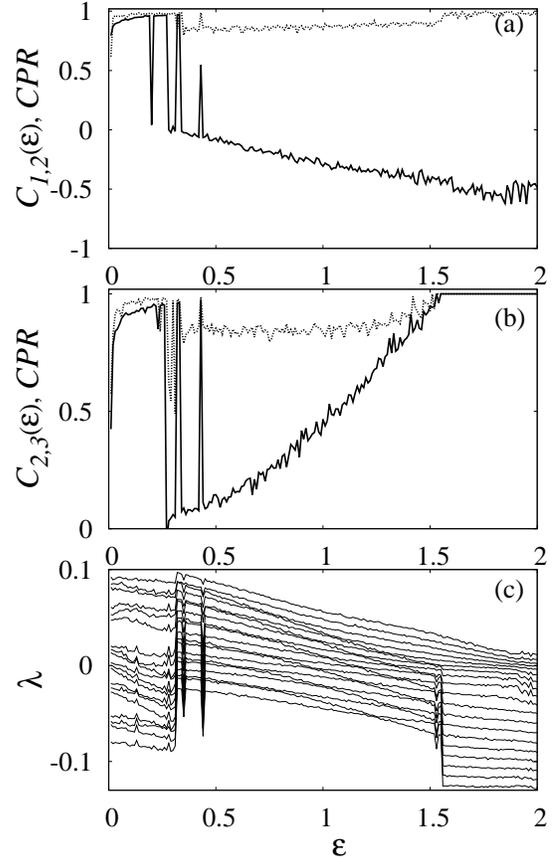}
\caption{\label{fig13} (a) Correlation coefficient $C_{1,2}(\varepsilon)$ and the
index CPR of the systems 1, 2, (b) correlation coefficient $C_{2,3}(\varepsilon)$
and the CPR of the two outer systems 2, 3 and (c) display of the transition of the 
largest LEs of the coupled systems as a function of the coupling strength in subsystem coupling configuration.}
\end{figure}

The transition from asynchronization to ZLS of the two outer systems are again characterized 
by calculating the correlation coefficient (with $\Delta t=0$ in Eq.~(\ref{eqn4a})),
the transition in LEs, while the phase coherence can be quantified by using the index CPR as a function of the coupling
strength $\varepsilon\in(0,2)$. Figures~\ref{fig13}(a) and \ref{fig13}(b) depict the 
correlation coefficient (continuous curve) and the index CPR (dotted curve) 
between the systems 1, 2  and 2, 3, respectively, as a function of $\varepsilon$ 
exhibiting the following regimes:

(i) For certain lower values of coupling strength ($\varepsilon<0.25$), CPR reaches 
the unit value while the correlation coefficients
$C_{1,2} (\varepsilon)$ and $C_{2,3} (\varepsilon)$ reach  values
near to unity (but not exactly 1), which indicate that all the three systems 
exhibit in-phase synchronization
(approximate complete synchronization) [Figs.~\ref{fig13}(a) and \ref{fig13}(b)].
This transition is also confirmed from the changes in the LEs of the coupled systems
where they become negative for the corresponding values of $\varepsilon$  [Fig.~\ref{fig13}(c)]. 

(ii) Beyound $\varepsilon>0.25$ there is a desynchronization transition, where the CPR and
correlation coefficient drop down well below unity and some of the positive LEs gradually
acquire negative values. 

(iii) If we continue to increase the coupling further, the CPR of the systems 1, 2 increases
 and reaches the unit value 
at $\varepsilon\approx 1.6$, whereas the correlation coefficient $C_{1,2}(\varepsilon)$
becomes negative as a function of $\varepsilon$, which also confirms 
the onset of inverse phase synchronization between systems 1 and 2  [Fig.~\ref{fig13}(a)].
But the correlation coefficient ($C_{2,3}(\varepsilon)$) and
the index CPR of the two outer systems 2, 3 increase as a 
function of the coupling strength and for $\varepsilon=1.55$, both the measures reach the unit value 
indicating the existence of persistent ZLS between the systems $2$ and $3$ [Fig.~\ref{fig13}(b)].

(iv) The above synchronization transition is further confirmed by the transition in the largest LEs of the
coupled systems (\ref{eqn5}). In Fig.~\ref{fig13}(c) we plot nineteen largest LEs (with nine positive LEs).
The first three LEs correspond to the largest LEs of the three independent systems. 
If we increase $\varepsilon$, the largest LEs of the outer systems (second and third largest LEs) 
become negative at $\varepsilon\approx1.55$ indicating
the occurrence of ZLS in the subsystem coupling configuration. 

(v) It is also to be noted that in the subsystem coupling 
configuration there always exist multiple positive LEs confirming the 
existence of ZLS in the hyperchaotic regime [Fig.~\ref{fig13}(c)].
In contrast to the case of mutual coupling, there exists no periodic regime
because one of the subsystems remains unaffected and ZLS occurs only in 
hyperchaotic regime. In addition, all the systems in 
the subsystem coupling configuration exhibit in-phase and approximate CS over a 
large range of $\varepsilon$ for lower coupling strengths compared to the 
occurence of a very few spikes of in-phase
and approximate CS in the mutual coupling configuration.

Finally, to carry out a linear stability analysis for the subsystem coupling configuration (\ref{eqn5}), 
again we consider the difference between the two outer systems $\Delta=x_{2}(t)-x_{3}(t)$. 
The error equation corresponding to the synchronization manifold,
$\dot{\Delta}=-(\alpha+\varepsilon)\Delta+\beta [A f^{\prime}(x_{2}(t-\tau))-B] \Delta_{\tau}$,
which is exactly same as Eq.~(\ref{eqn1a}) in Sec.~\ref{stability}. 
So the sufficient condition for the stability of ZLS is also the same as in Eq.~(\ref{eqn1e}).
Also from the numerical results we find that ZLS between the two outer systems occur 
for $\varepsilon=1.55$, which satisfies the condition $(\alpha+\varepsilon)>|\beta(A-B)|\approx (1.55>1.04)$.

\section{\label{sec:level5}Conclusion}
We have reported here the first experimental confirmation of ZLS in a system of three coupled identical
piecewise linear time-delayed electronic circuits via dynamical relaying with different 
coupling configurations, namely mutual and subsystem coupling configurations. 
From the obtained results, we have identified that when
there is a feedback between the central and at least one of the outer systems, 
occurrence of ZLS in the two outer systems takes place, while the central and outer 
systems exhibit IPS. In the above two cases, the central unit plays a key role to obtain ZLS. 
In the case of mutual coupling configuration ZLS occurs in both periodic
and hyperchaotic regimes, while in the subsystem coupling configuration it occurs only in the hyperchaotic regime.
We also find that for certain lower values of the coupling strength all the three systems 
exhibit in-phase synchronization and for large coupling (here $\varepsilon>1.5$) both outer
systems exhibit ZLS, while the central and outer systems are in IPS.
The results are experimentally confirmed from snapshots of the time evolution
and phase projections of the outer systems. The quality of synchronization is numerically
verified using the correlation coefficient and the similarity function. 
The transition to ZLS is characterized and confirmed from the changes in the largest LEs and 
the correlation coefficient of the coupled systems as a function of the coupling strength. 
Further, IPS is experimentally confirmed using snapshots of time series 
plots and the phase coherence is characterized both qualitatively and quantitatively by using the 
concept of localized sets and the index CPR, respectively. We have also analytically 
derived the stability condition to confirm the occurrence of ZLS using the
Krasovskii-Lyapunov theory. One may also add that even though we have 
considered only a system of three coupled oscillators, one can treat each of the 
oscillators as corresponding to one group of synchronized
oscillators. In fact, one can consider such groups of oscillators with
different topologies to explain neuronal information processing among different
parts of the brain, synchronization among groups of birds or animals in ecology
in spite of large spatial separation, etc. Work is in progress along these lines
which will be reported separately.

\section*{Acknowledgments}
The work of R.S., K.S. and M.L. has been supported by the Department of Science 
and Technology (DST), Government of India sponsored IRHPA research project.
M.L. has also been supported by a DST Ramanna project and a DAE Raja Ramanna Fellowship.
M.L. also acknowledges the support by the Alexander von Humboldt Foundation to visit PIK,
where the work was completed.
D.V.S. and J.K. acknowledge the support from EU under project No. 240763 PHOCUS(FP7-ICT-2009-C).


\end{document}